\documentclass[12pt,tightenlines,floats,aps,prd,twoside,onecolumn,english,
	superscriptaddress,showkeys,preprintnumbers,nofootinbib,showpacs,
	amssymb,amsmath,amsfonts]{revtex4}

\usepackage{babel}
\usepackage{amsmath}
\usepackage{amsfonts}
\usepackage{amssymb}
\usepackage{dsfont}

\begin{document}

\preprint{IGC-09/3-1}

\setcounter{page}{1}

\title{A possible topological interpretation of the Barbero--Immirzi parameter}

\author{\firstname{Simone} \surname{Mercuri}}
\email{mercuri@gravity.psu.edu}
\affiliation{Institute for Gravitation and the Cosmos, The Pennsylvania State University,\\ Physics Department, Whitmore Lab, University Park, PA 16802, USA}

\begin{abstract}
A possible topological interpretation of the Barbero--Immirzi parameter is proposed. Classically, by generalizing the Holst action to torsional spacetimes, we construct a precise analogy between the Barbero--Immirzi parameter and the $\theta$-angle of Yang--Mills gauge theories, where the role of the Pontryagin class is played by a well known topological term containing the Holst modification, the Nieh--Yan density. Quantum mechanically, the same analogy requires to study the large gauge sector of the theory. In particular, it is argued that the large sector of the gauge group can be correlated with the Nieh--Yan density, while the Barbero--Immirzi parameter plays the role of the free angular parameter of the large gauge transformations.
\end{abstract}

\date{March 12, 2009}

\pacs{04.20.Fy, 04.20.Gz}

\keywords{Barbero--Immirzi parameter, Large gauge transformations, Time gauge}

\maketitle

\section{Introduction}

In a recent paper \cite{Mer08}, the constraints of the Ashtekar--Barbero (AB) canonical formulation of gravity \cite{Ash86-87,Ash87-88,Bar95} have been obtained by suitably rescaling the heuristic quantum wave functional of the Einstein--Cartan (EC) theory by the exponential of the Nieh--Yan (NY) functional \cite{NieYan82}. In that construction, the so called Barbero--Immirzi (BI) parameter \cite{Imm97} played a role analogous to that of the $\theta$-angle in Yang--Mills gauge theories (see, e.g., \cite{Wei96} for a general description and \cite{Ash91} for a canonical approach), suggesting that its origin is correlated to the large component of the gauge group.

This interesting interpretation of the BI parameter was originally proposed by Gambini, Obregon and Pullin \cite{GamObrPul99}. Essentially, they pointed out the analogies existing between the appearance of the so-called $\theta$-angle in Yang--Mills gauge theories and the BI parameter in canonical quantum gravity. In fact, the BI parameter, $\beta$, shares many common features with the $\theta$-angle, in particular both do not affect the classical dynamics, nevertheless they produce striking effects in the quantum regime. But, while the theoretical interpretation of the $\theta$-angle of non-abelian gauge theories is clear and can be easily traced back to the non-trivial global structure of the gauge group \cite{Wei96,Pec98}, the situation regarding $\beta$ is more subtle.

\section{Ashtekar--Barbero formalism and the role of the Barbero--Immirzi parameter}
In order to introduce the reader into this argument, it is worth recalling some fundamental structures of canonical gravity. By introducing the Ashtekar self-dual $SL(2,\mathbb{C})$ connections in the framework of canonical General Relativity (GR), a Gauss constraint, which incorporates the generators of the local Lorentz boosts and rotations in a complex combination, appears besides the vectorial and scalar constraints, both connected with the diffeomorphisms gauge invariance of the theory. Simultaneously, the high non-linearity of the Arnowitt--Deser--Misner (ADM) canonical formulation of GR disappears: the new canonical constraints depend polynomially on the fundamental variables, both in vacuum and in the presence of matter \cite{AshRomTat89}. By using the Ashtekar formulation of GR, a background independent quantum theory of gravity was later formulated \cite{Ash91}. But the use of complex fundamental variables generates a serious difficulty connected with the implementation of the \emph{reality conditions} in the quantum theory, which are strictly necessary to ensure that the evolution of the system is real. This difficulty has not been overcome so far and, basically, it can be considered the technical motivation which led to the adoption of the real AB connections as fundamental variables, instead of the complex ones. The link existing between real and complex variables can be clarified by observing that both are obtainable from the ADM canonical pair via a contact transformation. In particular, a suitable canonical transformation allows to introduce a finite complex number, $\beta\neq 0$, namely the BI parameter, in the definition of the new variables, so that they correspond to the (anti)self-dual ones when $\beta=\pm\,i$ and to the real ones for any real value of $\beta$.

Geometrically, the main difference between these two sets of possible new variables for GR is the following: while the complex connections are the projection over the 3-space of the self-dual part of the Ricci spin connections, the real ones are non-trivially related to them, complicating their reconstruction \cite{Sam00}. In fact, the real $SU(2)$ valued connections contain only half of the necessary information for reconstructing the Lorentz valued connections of GR \cite{Thi01}, motivating also the necessity of fixing the temporal gauge in order to avoid the appearance of second class constraints.\footnote{The temporal gauge fixing consists in rotating the local basis by using a suitable Wigner boost so that, at every instant of ``time'', its zeroth component is parallel to the normal vector to the instantaneous Cauchy hypersurface $\Sigma^3_t$. This condition reduces the local $SO(3,1)$ gauge group to the subgroup of spatial rotations, $SO(3)$, by fixing the boost component of the Lorentz symmetry.} By fixing the temporal gauge, the accessible part of the phase space is determined by first class constraints only \cite{BarSa00} and the system can be quantized through the Dirac procedure. The result is a non-perturbative background independent quantum theory of gravity called \emph{Loop Quantum Gravity} (LQG) \cite{Thi01,Rov04,AshLew04,AshRovSmo92}.\footnote{LQG besides providing interesting physical predictions as the quantization of areas and volumes \cite{Rov04} (see also \cite{DitThi07,Rov07-1}), has been able to cure the inevitable singular behavior of classical GR in symmetric spacetimes \cite{Mod04,AshPawSin06,AshBoj06}. Furthermore, the recently obtained results about the graviton propagator have strengthened the physical content of the theory, providing new insights into its non-singular behavior \cite{Rov06,BiaModRov06,ChrLivSpe07}.}

Since the BI parameter has been introduced via a canonical transformation, one can naively believe that different values of $\beta$ correspond to unitary equivalent quantum theories. Strangely enough, this is not the case. In fact, $\beta$ enters in the spectrum of the main geometrical observables of the theory, e.g. the spectra of the area and volume operators, revealing that a one parameter family of non-equivalent quantum theories exists.
As argued by Rovelli and Thiemann \cite{RovThi98}, two dynamically equivalent $SO(3)$-valued connections exist and, as a consequence, an ambiguity appears in the theory, which is essentially expressed by the presence of the BI parameter.

Immirzi suggested that the appearance of the BI parameter in the quantum theory was a consequence of the temporal gauge fixing \cite{Imm97}, so that it would have disappeared in a fully Lorentz covariant theory. But this expectation was not completely confirmed by the so-called \emph{Covariant Loop Quantum Gravity} (CLQG), which is a fully Lorentz covariant quantum theory of gravity, constructed \emph{a l\`a} Dirac relaxing the time gauge condition \cite{Ale02}.\footnote{It is worth remarking that the complicated form of the Dirac parentheses, necessary to solve the second class constraints, prevents the fully Lorentz covariant theory from being rigorously formalized.} This approach, in fact, revealed a correlation between the choice of the fundamental variables and the appearance of the BI ambiguity in the quantum theory. In other words, in CLQG different choices of the fundamental variables are possible. In particular, for a geometrically well motivated specific choice of variables the resulting area spectrum no longer depends on the BI parameter \cite{AleVas01}. But, choosing different fundamental variables considered as a direct generalization of the AB connections, the resulting area spectrum turns out to depend on the BI ambiguity \cite{AleLiv03}, reproducing the result of the gauge fixed theory (see also the interesting paper \cite{CiaMon09}). 

These hints, together with the suggestion of Gambini, Obregon and Pullin \cite{GamObrPul99} and the observation described above about the functional relation between the EC and the AB constraints, lead to the idea that the structure of the large gauge sector of the theory can encompass the existence of the BI ambiguity. This could clarify its supposed topological origin and, as a consequence, the existence of non-unitary equivalent quantum theories associated to different values of $\beta$.

If this is the case, it must exist a classical framework where the analogy between the BI parameter and the $\theta$-angle of Yang--Mills gauge theories can be completed. In the pure gravitational case, in fact, the argument proposed fails to be completely convincing. The BI parameter appears in the action as a multiplicative constant in front of the so-called Holst modification \cite{Hol96},
\begin{align}\label{holst action}
S_{\rm Hol}\left[e,\omega\right]=\frac{1}{16\pi G\beta}\int e_a\wedge e_b\wedge R^{a b}\,,
\end{align}
which, in fact, is not a topological density. It does not reduce to a total divergence, rather it is an on-shell identically vanishing term. But the action (\ref{holst action}) can be further generalized to include in the picture also the interesting case of torsional spacetimes. In particular, in \cite{Mer06,Mer06P}, by introducing spinor matter fields, we paved the way to complete the Holst picture; specifically, we demonstrated that the presence of spinors can generate the necessary torsion contribution to generalize the Holst modification and construct a topological term. In other words, by using a non-minimal coupling between spinors and gravity,\footnote{See \cite{Kau08} for the extension to supergravity theories.} we,
indirectly, demonstrated that the EC action can be generalized without modifying the classical dynamics by adding the NY topological density \cite{NieYan82}, i.e. 
\begin{align}\label{new action for gravity}
\nonumber S&_{\rm Grav}=S_{\rm HP}\left[e,\omega\right]+S_{\rm D}\left[e,\omega,\psi,\overline{\psi}\right]+S_{\rm NY}\left[e,\omega\right]
=-\frac{1}{16\pi G}\int e_a\wedge e_b\wedge\star R^{ab}
\\
&+\frac{i}{2}\int\star e_a\wedge\left(\overline{\psi}\gamma^a D\psi-\overline{D\psi}\gamma^a\psi+\frac{i}{2}\,m e^a\overline{\psi}\psi\right)
+\frac{1}{16\pi G\beta}\int\left(e_a\wedge e_b\wedge R^{a b}-T^a\wedge T_a\right)\,.
\end{align}
By remembering the definition of the torsion 2-form $T^a=d e^a+\omega^a_{\ b}\wedge e^b$, the NY term can be easily rewritten as a total divergence, i.e.
\begin{equation}\label{Nieh--Yan}
\int\left(T^a\wedge T_a-e_a\wedge e_b\wedge R^{a b}\right)=\int d\left(e_a\wedge T^a\right)\,.
\end{equation}
The modification is now a true topological term related to the Pontryagin classes \cite{ChaZan97}, so that new interesting insights can be provided on the physical origin of the BI parameter. Moreover, this generalization is quite natural \cite{DatKauSen09,Mer09} and motivated the approach we presented in \cite{Mer08} as well as other recent works \cite{CalMer09,MerTav09}. But the structure of the large gauge group, which is supposed to be at the base of the proposed interpretation of the BI parameter, is still missing.

This is exactly the problem we will face in the remaining part of this paper. Specifically, we propose a possible link between the large gauge sector of the temporal gauge fixed gravitational theory and the NY density, also making a useful comparison with the well known example of $SU(N)$ Yang-Mills gauge theories.

\section{Large gauge transformations in Yang-Mills gauge theories}\label{Sec2} 
Let the $SU(N)$ valued connection $A_{\alpha}=\sum_I A_{\alpha}^I\lambda^I$ and its associated electric field $E^{\gamma}=\sum_{K}E^{\gamma}_K\lambda^K$ (where $I,J,K,\cdots$ are internal indexes running on $1,2,\cdots,N^2-1$) be a couple of conjugate variables in the framework of a canonical formulation of Yang-Mills gauge theories. The evolution of the system is limited to a restricted region of the phase space by the first class Gauss constraint, expressed by the following weak equation:
\begin{equation}\label{Gauss Y-M}
G_I:=D_{\alpha}E^{\alpha}_I=\partial_{\alpha}E^{\alpha}_I+f_{I J}^{\ \ K}A^{J}_{\alpha}E^{\alpha}_K\approx 0\,.
\end{equation}
According to the Dirac quantization procedure \cite{Dir64,HenTei92}, the state functional describing the quantum physical system must satisfy the Gauss constraint (\ref{Gauss Y-M}), namely we have to require that
\begin{equation}
\widehat{G}_I\Phi(A)=-i D_{\alpha}\frac{\delta}{\delta A_{\alpha}^I}\Phi(A)=0\,,
\end{equation}
where the usual quantum representation of the operators has been assumed. 

The Gauss constraint in Eq.(\ref{Gauss Y-M}) formalizes the request of gauge invariance of the quantum state describing the physical system, namely it is equivalent to requiring that the state functional be invariant under the small component of the gauge group $G=SU(N)$, as can be easily realized. Since the global structure of the gauge group is non-trivial, in view of quantization, it is particularly interesting to study the behavior of the state functional under the large gauge transformations. It, in fact, can produce striking effects in the non-perturbative theory, as, e.g., $P$ and $CP$ violations, physically motivating this extension of the theory.

In this respect, let $\widehat{\mathcal{G}}$ be the generator of the large gauge transformations, acting on the state functional $\Phi(A)$. Considering that the Hamiltonian operator, $\widehat{\mathcal{H}}$, is invariant under the full gauge group (or, more formally, it commutes with the operator $\widehat{\mathcal{G}}$), we can construct a set of eigenstates for the quantum theory by diagonalizing simultaneously $\widehat{\mathcal{H}}$ and $\widehat{\mathcal{G}}$. In other words, the following equation
\begin{equation}\label{large operator}
\widehat{\mathcal{G}}\Phi_w(A)=\Phi_w(A^{g})=e^{i\theta w}\Phi_w(A)\,,
\quad\text{where}\quad
A^{g}=g Ag^{-1}+g dg^{-1}\,,
\end{equation}
is a super-selection rule for the states of the theory, which are now labeled by the \emph{winding number} $w=w(g)$, according to their behavior under the action of the large gauge transformation operator. The constant $\theta$ introduced in Eq. (\ref{large operator}) is an angular parameter, which indicates how much the state functional ``rotates'' under the action of the large gauge transformations operator. Specifically, it represents a quantization ambiguity connected with the non-trivial global structure of the gauge group.

Eq.(\ref{large operator}) implies that the wave functionals have to satisfy suitable $\theta$-dependent boundary conditions passing from one ``slab'' to the next in the configuration space; or, a fully gauge invariant state functional can be constructed, transferring the $\theta$ dependence in the momentum operator. In this respect, we recall that the so-called \emph{Chern-Simons functional},
\begin{equation}
\mathcal{Y}(A)=\frac{1}{8\pi^2}\int{\rm tr}\left(F\wedge A-\frac{1}{3}A\wedge A\wedge A\right)\,,
\end{equation}
is characterized by the following remarkable property: 
\begin{equation}
\mathcal{Y}\left(A^{g}\right)=\mathcal{Y}\left(A\right)+w(g)\,.
\end{equation}
This directly implies that the new state functional, 
\begin{equation}\label{rescaling Y-M}
\Phi^{\prime}(A)=e^{-i\theta\mathcal{Y}(A)}\Phi_w(A)\,,
\end{equation}
will be invariant under the full gauge group, as can be easily demonstrated.

So, by using the rescaling in Eq.(\ref{rescaling Y-M}), we have obtained a new fully gauge invariant quantum state functional, at the price of modifying the momentum operator. In other words, the $\theta$-dependence has been transferred from the boundary conditions to the momentum operator, which becomes:
\begin{align}
E^{\prime}_{\alpha}\Phi^{\prime}(A)=e^{-i\theta\mathcal{Y}(A)}E_{\alpha}e^{i\theta\mathcal{Y}(A)}\Phi^{\prime}(A)=-i\left[\frac{\delta}{\delta A^{\alpha}}-\frac{i\theta}{8\pi^2}\,\epsilon_{\alpha}^{\ \beta\gamma}F_{\beta\gamma}\right]\Phi^{\prime}(A)\,.
\end{align} 
The above modification in the conjugated momentum reflects on the Hamiltonian operator, i.e.
\begin{align}
H^{\prime}=\int d^3x\,{\rm tr}\left[ \frac{1}{2}\,\left(E_{\alpha}-\frac{\theta}{8\pi^2}\,\epsilon_{\alpha}^{\ \beta\gamma}F_{\beta\gamma}\right)^2+\frac{1}{4}\,F_{\alpha\beta}F^{\alpha\beta}\right]\,,
\end{align}
generating a pseudo-vectorial term which prevents the new Hamiltonian $H^{\prime}$ from being invariant under the CP discrete symmetry. 

The new Hamiltonian corresponds to a topological modification of the classical action, consisting in the presence of an additional term belonging to the Pontryagin class, i.e.
\begin{equation}\label{Y-M top}
S_{new}(A)=-\frac{1}{4}\int{\rm tr} \star F\wedge F+\frac{\theta}{8\pi^2}\int{\rm tr} F\wedge F\,.
\end{equation}
The $\theta$ parameter appears as a multiplicative constant in front of the modification. It is worth mentioning the fact that the new term does not affect the classical equations of motion, but modifies the vacuum to vacuum amplitude in the path-integral formulation of the quantum theory. In other words, it allows to take into account possible tunneling phenomena between vacua characterized by different winding numbers, violating the CP discrete symmetry. 

\section{Large gauge transformations in partially gauge fixed gravity}
Many of the topological aspects of Yang--Mills gauge theories seems to have an analogous  counterpart in the gravitational theory. On one hand, as previously said, the action for gravity can be generalized through the addition of the NY topological term in such a way it resembles to the action (\ref{Y-M top}), where the BI parameter plays a role analogous to the $\theta$-angle \cite{Mer06,Mer08,DatKauSen09}. On the other hand, we have recently demonstrated \cite{Mer08} that, in fact, the constraints associated to the modified action (\ref{new action for gravity}) can be obtained from the constraints of the EC canonical theory by rescaling the heuristic wave functional as in Eq. (\ref{rescaling Y-M}). Specifically, the role of the Chern--Simons is played by the NY functional
\begin{equation}\label{NY functional}
\mathcal{Y}\left[e,\omega\right]=\int e_i\wedge T^i\,,
\end{equation}
($i,j,k\dots$ are internal indexes), while the associated angular parameter turns out to be exactly the BI parameter. The rescaling, as described previously, implies a redefinition of the momentum operator, which, once put back in the original constraints of the EC theory, exactly leads to the constraints obtainable starting from action (\ref{new action for gravity}) \cite{Mer08} (see \cite{DatKauSen09} for the classical canonical formulation of the modified gravitational theory with fermions).

The missing point in this construction is the relation existing between the NY and the large gauge sector of the theory, in analogy with the requirement of invariance under the large sector of the $SU(N)$ gauge group pertaining to the case of Yang--Mills gauge theories.

In order to understand this point, we have to take into account the fact that the EC as well as the AB first class constraints are extracted from the fully covariant theory after having fixed the temporal gauge. This fixes the zeroth component of the local basis, $e^0$, in such a way that it remains parallel to the normal vector, $n$, along the evolution and, simultaneously, reduces the gauge group from $SO(3,1)$ to $SO(3)$ (see footnote 1).\footnote{We could choose a different gauge condition by acting on the local basis with a general Wigner boost of parameters $\chi^k$. This, in fact, would still fix the zeroth component of the local basis reducing the local gauge group to $SO(3)$, though, it does not require any specific condition on $e^0$ in contrast with the widely used temporal gauge. The temporal gauge is usually preferred, because it considerably simplifies the canonical analysis \cite{NelTei78}, nevertheless, every admissible gauge choice must be physically equivalent.}

Therefore, once the gauge has been partially fixed, the local symmetry group reduces to the group of spatial rotations, $SO(3)$, so that one is immediately induced to think that the large gauge sector is merely related to the non-trivial global structure of $SO(3)$. But, physically, also the action of the $T$ discrete operator, which acts on the zeroth component of the local basis by flipping its orientation with respect to the normal vector, represents a large gauge transformation. As a consequence the full large gauge group is $\mathcal{G}=SO(3)\times \mathbb{Z}_2\simeq S^3$. Namely, it consists of two copies of $SO(3)$, correlated with the two orientations of the zeroth component of the local basis.\footnote{It is worth noting the difference existing by fixing the temporal gauge in such a way that $e^0$ is anti-parallel to $n$ (acting with a specific Wigner boost on the local basis) and considering the action of the operator $T$, which is a large gauge transformation and changes the right-left hand character of the local basis.} In particular, recalling that $\Pi_3(S^3)=\mathbb{Z}$, the disconnected components of the large gauge group are labeled by an integer, which is the winding number of the $SU(2)\simeq S^3$ group. 

Analogously to the Yang--Mills case discussed in the previous section, in order to construct states which are invariant under the large gauge group, a rescaling of the state functional of the canonically quantized EC theory by the exponential of a suitable Chern--Simons-like  functional is required. We claim that the natural functional able to incorporate the global properties of the large gauge group of gravity is the one in Eq.(\ref{NY functional}). In order to motivate our claim, we digress on a simple geometrical construction, initially illustrating the situation in a lower dimensional case. We follow the procedure described in \cite{Wis06}, which will be useful also in the more relevant case.

Consider a 2-dimensional manifold, the compact group $SO(2)$ acts naturally on the tangent bundle, locally rotating the local basis on the two dimensional planes $\mathbb{R}^2$. Now, imagine to construct a new bundle, by extracting a point from the local planes $\mathbb{R}^2$ and compactifying it to a sphere $S^3$, in other words, the new bundle is made up of the tangent spheres obtained by one-point compactification of $\mathbb{R}^2$. On the new bundle, we can consider the natural action of the compact group $SO(3)$, which represents an enlargement of the original local group $SO(2)$. Enlarging the gauge group from $SO(2)$ to $SO(3)$ is equivalent to considering the parallel transport of spheres instead of the parallel transport of planes \cite{Wis06}. The movements of the spheres can be factorized in an $SO(2)$ rotation which maintains fixed the tangent point and a $S^1$ translation of the tangent point itself. So that the connection involved in the parallel transport of the sphere is expected to be of the MacDowell--Mansouri form \cite{MDoMan77}.

In the case under consideration of gauged fixed gravity, an analogous construction will allow us to show some interesting topological properties of the NY functional. The temporal gauge fixing reduces the local group from $SO(3,1)$ to the subgroup $SO(3)$. Namely there is a natural action of the group $SO(3)$ on the vectors of the local spatial basis, $e^i$. The freedom in choosing the orientation of the zeroth component of the 4-dimensional local basis with respect to the normal vector corresponds to an enlargement of the large gauge group from $SO(3)$ to $SO(3)\times\mathbb{Z}_2$, as argued before. In order to characterize its topological structure using the ingredients of canonical gravity, we can initially construct a connection for the $SO(4)$ group and then considering the Chern--Simons functional associated to the quotient $SO(4)/SO(3)$. In order to avoid any confusion, we want to stress that enlarging the gauge group to $SO(4)$ is just a mathematical tool without any particular physical meaning. We consider this procedure useful for clarifying the interpretation of the NY functional as the Chern--Simons of the large gauge group $\mathcal{G}=SO(3)\times\mathbb{Z}_2$, which is rewritable as a quotient, i.e. $\mathcal{G}=SO(4)/SO(3)$. 

As in the lower dimensional case, the connection of the $SO(4)$ can be written in the MacDowell--Mansouri form, i.e.
\begin{equation}
\Omega^{A B}=\left(\begin{array}{cc}
\Gamma^{i j} & \frac{1}{\ell}\,e^i 
	\\[12pt]
	-\frac{1}{\ell}\,e^j & 0
\end{array}
\right)\,,
\end{equation}
where $A,B,C,\dots$ are indexes valued on $SO(4)$, while $i,j,k,\dots$ are valued on $SO(3)$. The constant $\ell$ has the dimension of a length and can be associated with the radius of the spheres obtained compactifying the tangent planes. By using the above connection it is easy to demonstrate that the following relation holds:
\begin{align}\label{ste}
\nonumber \mathcal{Y}\left[\Omega\right]&=F^{A B}\wedge\Omega_{A B}+\frac{1}{3}\Omega^A_{\ B}\wedge\Omega^B_{\ C}\wedge\Omega^C_{\ A}
\\
&=R^{i j}\wedge\Gamma_{i j}+\frac{1}{3}\Gamma^i_{\ j}\wedge\Gamma^j_{\ k}\wedge\Gamma^k_{\ i}-\frac{2}{\ell^2}T^i\wedge e_i=\mathcal{Y}\left[\omega\right]-\frac{2}{\ell^2}\mathcal{Y}\left[e,\omega\right]\,,
\end{align}
where $F^{A B}$ is the curvature 2-form associated with the connection $\Omega^{A B}$, while $R^{i j}$ is associated with the 3-dimensional connection $\Gamma^{i j}$.

Now, considering that $\mathcal{G}=SO(4)/SO(3)$, we can construct a Chern--Simons functional for the large gauge group of gauge fixed gravity as the difference between $\mathcal{Y}\left[\Omega\right]$ and $\mathcal{Y}\left[\omega\right]$, but this is exactly the Nieh--Yan functional as can be immediately understood looking at Eq. (\ref{ste}). Namely we have: 
\begin{equation}
\mathcal{Y}\left[e,\omega\right]=\frac{\ell^2}{2}\left(\mathcal{Y}\left[\omega\right]-\mathcal{Y}\left[\Omega\right]\right)\,,
\end{equation}
so that we can conclude that the NY functional is correlated with the large gauge group of gauge fixed gravity, $\mathcal{G}=SO(3)\times \mathbb{Z}_2=SO(4)/SO(3)$, as initially argued. Finally, a new state functional, fully invariant under the large gauge group, can be obtained by rescaling the original state functional of the EC theory by the NY functional: as previously proved in \cite{Mer08}, the new state functional satisfies the AB constraints for GR, revealing the topological origin of the BI parameter.

\section{Conclusions}
By generalizing the Holst action to torsional spacetimes, we have argued that the BI parameter can have a topological origin analogous to that of the $\theta$-angle of Yang--Mills gauge theories. But, while the appearance of the $\theta$ parameter can be traced back to the non-trivial global structure of the gauge group, the situation regarding $\beta$ was less clear. Here we have described how the BI parameter is connected through the NY functional to the large gauge sector of canonical gravity. In particular, fictitiously enlarging the local group to $SO(4)$ and then considering that the large gauge group of temporal gauge fixed gravity is $\mathcal{G}=SO(4)/SO(3)$, we also clarified the role of the NY functional in analogy to that of the Chern--Simons in Yang--Mills gauge theory. 

It is worth noting that this argument suggests that the AB canonical formulation of gravity represents a non-trivial extension of the EC theory. Specifically, the presence in the action of the NY invariant, introduces into the theory also information about the global structure of the local gauge group. Furthermore, the argument described in this paper indicates that the states satisfying the AB quantum constraints should not be affected by the action of the operator $T$, which reverses the orientation of the zeroth component of the local basis; further details will be given elsewhere.

\begin{acknowledgments}
The author would like to thank S. Alexandrov for an interesting and clarifying mail exchange and A. Randono for discussion. This research was supported in part by NSF grant PHY0854743, The George A. and Margaret M. Downsbrough Endowment and the Eberly research funds of Penn State.
\end{acknowledgments}

\end{document}